# Spin-orbit torques driven by the interface-generated spin currents

Q. B. Liu[1,2], K. K. Meng[1]*, S. Q. Zheng, Y. C. Wu[1], J. Miao[1], X. G. Xu[1], Y. Jiang[1]*

[1]*Beijing Advanced Innovation Center for Materials Genome Engineering, University of Science and Technology Beijing, Beijing 100083, China*

[2]*Applied and Engineering Physics, Cornell University, Ithaca, NY 14853, USA*

**Abstract:** The spin currents generated by spin-orbit coupling (SOC) in the nonmagnetic metal layer or at the interface with broken inversion symmetry are of particular interest and importance. Here, we have explored the spin current generation mechanisms through the spin-orbit torques (SOTs) measurements in the Ru/Fe heterostructures with weak perpendicular magnetic anisotropy (PMA). Although the spin Hall angle (SHA) of Ru is smaller than that in Pt, Ta or W, reversible SOT in Ru/Fe heterostructures can still be realized. Through non-adiabatic harmonic Hall voltage measurements and macrospin simulation, the effective SHA in Ru/Fe heterostructures is compared with Pt. Moreover, we also explore that the spin current driven by interface strongly depends on the electrical conductivities. Our results suggest a new method for efficiently generating finite spin currents in ferromagnet/nonmagnetic metal bilayers, which establishes new opportunities for fundamental study of spin dynamics and transport in ferromagnetic systems.

*kkmeng@ustb.edu.cn

**yjiang@ustb.edu.cn

The current-induced magnetization dynamics mediated by spin-orbit coupling (SOC) offers a wide spectrum of opportunities to add the spin degree of freedom to conventional charge-based microelectronic devices or to completely replace charge with spin functionalities [1-4]. Over the past few years, theoretical and experimental investigations of spintronics have offered means to read and write information in nanomagnets by a new family of spin torques. These torques originate from the transfer of orbital angular momentum from the lattice to the spin system and are called spin-orbit torques (SOTs) in order to underline their direct link to the SOC. In general, the SOT relies on the conversion of electrical current to spin, and there are two main conversion mechanisms: the spin Hall effect (SHE) and the inverse spin galvanic effect (ISGE) [5-7]. The SHE is a collection of SOC phenomena in which electrical currents can generate transverse spin currents and vice versa in the nonmagnetic heavy metal (HM) layer. It originates from three distinct microscopic mechanisms: the skew, the side jump, and the intrinsic mechanisms. The mechanisms are caused by coherent band mixing effects induced by the external electric field and the disorder potential. The spin current propagates toward the interface, where it is absorbed in the form of a magnetization torque in the adjacent ferromagnet (FM), which shares with the spin transfer torques the basic concept of the angular momentum transfer from a carrier spin current to magnetization torque. From this perspective, the dominant component of the SHE-SOT in this picture is damping-like and takes the form $\tau_D \sim \boldsymbol{m}\times(\boldsymbol{m}\times\boldsymbol{\sigma})$, where $\boldsymbol{m}$ and $\boldsymbol{\sigma}$ are unit vectors of the magnetization in the FM and the non-equilibrium spin polarization direction in the HM [8]. However, the spin current also yields a nonequilibrium spin density at the edges of the HM where the inversion symmetry is broken. This implies an alternative picture of the SHE-SOT caused by the nonequilibrium spin density at the HM/FM interface. Correspondingly, the SHE can also be expected to contribute to the field-like torques $\tau_F \sim m\times\sigma$. In the ISGE, a charge current flowing parallel to an interface with broken inversion symmetry generates a spin density due to SOC, which in turn exerts a torque on the magnetization of an adjacent FM via the exchange coupling [9]. Similarly, the ISGE mechanism can yield both damping-like and field-like SOT terms. These two types of torques equivalently yield damping-like effective field $H_D$ and field-like effective field $H_F$, respectively [10].

The interfaces in HM/FM heterostructures could induce an extraordinary range of phenomena attributed to finite thickness, symmetry breaking, atomic and electronic reconstructions, and interaction effects [11-13]. These effects are intrinsic in that they would occur even at ideal interfaces between defect-free materials, which do not rely on imperfections for their existence. However, ideal interfaces between defect-free materials do not generally exist. This limit may be approached when kinetic limitations dominate thermodynamics, but in most cases nonidealities are inherent. A detailed appreciation of the physics and chemistry of interfaces is often required to understand interfacial magnetic behavior [14]. Notably, defects and disorder are not universally deleterious, but can induce, and even control, novel phenomena. There are many examples where defects are even more significant, being the origin of the spin current [15, 16]. On the other hand, the usage of some light metals (LM) with weaker SOC such as V, Cr and Mo have also been proved to generate substantial spin current in the LM/FM heterostructures [17-19]. Therefore, the SOT is highly interface-sensitive and the interfacial contribution to SOT requires reexamination using three-dimensional model instead of only regarding as a typical two-dimensional interface. Recent theoretical results have shown that the carriers can exhibit a net spin polarization and carry a net spin current near interfaces with SOC by using three-dimensional solutions of the spin-dependent Boltzmann equation [20]. It indicates that the interface-generated spin currents, even though under the weak interface effective spin-orbit field, would be comparable to the SHE of Pt. It has been experimentally demonstrated through the other-layer-generated SOT in magnetic trilayers [21]. However, the complex magnetic multilayers impede further detecting the physical mechanisms of spin current generation since only the bottom interface is regarded as spin current source, without involving the discussion of the top interface.

In this work, we have investigated the mechanisms of spin current generation in the Ru/Fe heterostructures, which show weak perpendicular magnetic anisotropy (PMA) and thermal stability at room temperature. It gives an effective and convenient platform to study the contribution of interfacial spin-orbit scattering, which is different from the complicated multilayered systems [22]. Although the SHA of Ru is smaller as compared with that in Pt, Ta or W, the current-induced SOT can still be realized. Through non-adiabatic harmonic Hall voltage measurements and macrospin simulation, the effective SOT fields are found to be dramatically depended on the Ru/Fe interfaces and comparable with that generated by SHE of Pt. The non-negligible interface-generated spin currents are considered to be widely presented in HM/FM bilayers and closely related with electrical conductivities.

For our experiments, samples of nonmagnetic metals (NMs) (Pt, Ru, Cu)/FMs (Fe, Co, CoFeB)/$HfO_2$ structures were deposited at room temperature on oxidized silicon wafers using magnetron sputtering, while $HfO_2$ was used as an insulating capping layer. The thickness of NM is 3 nm, FM 1 nm and $HfO_2$ 3 nm, respectively. Before the deposition, the electron beam lithography and Ar ion milling were used to pattern Hall bars, and a lift-off process was used to make contact electrodes. The size of all the Hall bars is 20 μm×120 μm, in which the long stripe is along the direction X and the short one is along the direction *Y*. The magnetic properties of the unpatterned samples were measured using a vibrating sample magnetometer (VSM). A sinusoidal AC current $I_{ac}$ was applied in the direction *X* to exert periodic SOT on the magnetization with a fixed frequency and different current amplitudes. The first $V_\omega$ and second $V_{2\omega}$ harmonic anomalous Hall voltages were measured simultaneously as a function of the magnetic field using two lock-in amplifier systems. The SOT-induced magnetization switching was measured by applying a pulsed current in the direction *X* with a width of 50 μs, and the resistance was measured after a 16 μs delay under an external magnetic field $H_X$ along the direction *X*.

Figure 1(a) shows the magnetic field ***H*** dependence of the out-of-plane magnetization ***M*** of Pt/Fe films, which indicates an in-plane magnetic anisotropy. To measure the current-induced torques on the magnetic layer (which can arise from either spin currents or an Oersted field), we used the second-harmonic Hall technique in which a low frequency (60 Hz) alternating current is applied to the device and the induced Hall voltages were measured at the second harmonic frequency. For all samples, we have applied a 5 mA signal, so that the current density within a given material layer is approximately the same for different samples. In principle, for an in-plane magnetic anisotropic film, the second-harmonic Hall technique can give the information of both the in-plane and out-of-plane components of current-induced torques, but one must be careful to distinguish the spin-torque signals arising from the artifacts associated with thermoelectric effects [23]. In-plane torques correspond to out-of plane effective magnetic fields (by the right-hand rule), so they tend to pull an in-plane magnetization slightly out of plane, giving a second harmonic Hall voltage signal on account of mixing an oscillating anomalous Hall resistance and the oscillating current. According to the theory of Hayashi *et al*, for the in-plane magnetized device, whose easy axis is along the *X* axis and $H_K < 0$, the out-of-plane field sweep gives information of the field-like term whereas the in-plane transverse field sweep provides that of the damping-like term when $|H_K| \gg |H_A|$ [24]. This is due to the larger contribution of the planar Hall effect (PHE) on the second harmonic signal ($V_{2\omega}$) over that of the anomalous Hall effect (AHE) [24]. We swept the magnetic field that was perpendicular or parallel to the current flow direction with a stably tilted field ($\theta_H = 5°$) at room temperature. As shown in Fig. 1(b), the first harmonic signal ($V_\omega$) with a small tilted angle mainly respects to the AHE. The $V_{2\omega}$ shows a positive peak at a positive magnetic field and a negative peak at a negative magnetic field with the magnetic field sweeping parallel and perpendicular to the current flow directions as shown in Figs. 1(c) and (d) respectively. Therefore, if the polarization of spin current is along direction Y, the harmonic signals $V_\omega$ and $V_{2\omega}$ can be simplified as [24]:

$$V_\omega = \frac{1}{2} \Delta R_A \cos\theta_0 \Delta I, \quad (1)$$

$$V_{2\omega} = -\frac{1}{2} \Delta R_A \Delta I [\frac{1}{2} \frac{\sin\theta_0 \cos\varphi_H}{H_K \cos 2\theta_0 + H \cos(\theta_H - \theta_0)} H_D + \xi \frac{\sin\theta_0 \cos\varphi_H}{-H_A \sin\theta_0 + H \sin\theta_H} H_F], \quad (2)$$

where $H_D$ and $H_F$ corresponds to the damping-like and the field-like field, respectively. $\xi = \Delta R_{PHE}/\Delta R_{AHE}$, $\theta_{0(H)}$ and $\phi_{0(H)}$ are the polar and azimuthal angles of the magnetization *M* (external magnetic field). Figs. 1(e) and (f) show the harmonic Hall data with a fitting according to the theory, which is inconsistent with the measured data [24]. Meanwhile, the $V_{2\omega}$ signals with the magnetic field sweeping with a stably tilted out of plane ($\theta_H = 5°$ and $\varphi_H = 90°$) should not stem from a conventional SOT due to the SHE, because the conventional spin transfer torque is zero for magnetization oriented perpendicular to the current flow. The anomalous Nernst effect (ANE) and spin Seebeck effect (SSE) can also be excluded, since the two mechanisms will induce a voltage that is proportional to $j^2 \cdot (\boldsymbol{m}_{Fe} \times \nabla T)$ and the thermoelectric signals would arrive at saturation with increasing magnetic field. On the contrary, the $V_{2\omega}$ is approximating to be zero at the saturation magnetic field [25]. Therefore, in the Pt/Fe heterostructures, the spin polarization direction should not always point along the $Y$ direction. The origin of spin current generation in Pt/Fe heterostructures needs to be further clarified which is different from the conventional SHE and ISGE.

Ru has been traditionally treated as an important material for the synthetic antiferromagnet structures, which has been widely used in magnetic sensors and recording media [26]. In this work, the Ru/Fe heterostructures are chosen to discuss the origin of the spin current in NM/HM heterostructures, in which the Ru layer has weak SOC and similar electrical conductivity with Pt. According to the magnetic field $H$ dependence of the out-of-plane magnetization $M$ and anomalous hall resistance $R_H$ with varying temperature as shown in Fig. 2 (a) and (b), we found the Ru/Fe heterostructures has weak PMA in a small field range (see details in supplementary S2). Then, the current-induced magnetization switching behaviors were measured with applying a non-zero in-plane field $H_X$ at room temperature. As shown in Fig. 3 (a), the switching direction changes from anticlockwise to clockwise as $H_X$ is changing from negative to positive and the fully magnetization switching is observed when $H_X$ is 150 Oe. The critical switching current $I_c$ is 80 mA, corresponding to the switching current density $J_c = 1 \times 10^8$ A/cm$^2$ [27] and the values versus different in-plane bias fields $H_X$ are shown in Fig. 3(b). The generated spin currents in the Ru/Fe heterostructures are comparable to measured spin current in Pt/Co/Pt system in our preview experiment even though the SOC in Ru is much smaller than Pt [28]. Note that the measured $J_c$ as shown in Fig. 3(b) initially decreased and finally increased at the applied $H_X$~300 Oe. To understand the slight change of damping-like

SOT efficiency under various $H_X$, the function of the (Dzyaloshinskii-Moriya interaction) DMI at the HM/FM interface needs to be considered [29]. If the applied $|H_X|$ is much greater than $|H_{DMI}|$, the magnetic moments in the film would become more single-domain-like, and thus, the SOT-driven or SOT-assisted domain nucleation/propagation is not as efficient as the proper $H_X$ case. This slight decrease of dampling-like SOT efficiency for $|H_X| \gg |H_{DMI}|$ has also been proved in the recent work [29]. To quantitatively determine the strength of the spin-orbit effective fields in the samples, the harmonic measurements with sweeping an in-plane magnetic field parallel or perpendicular to the current directions were carried out. The $V_\omega$ and $V_{2\omega}$ against larger $H_X$ were shown in supplementary S3. The small field harmonic measurements by applying a sinusoidal AC current with the amplitude of $8\times10^6 A/cm^2$ at 300 K is taken as an example as shown in Fig. 3 (c), in which the black and red lines are corresponding with the out-of-plane magnetization components $M_Z > 0$ and $M_Z < 0$, respectively. Before the harmonic measurements, we have applied a large out-of-plane external field to saturate the multilayers magnetization, which remains saturated after the field was turned off. Therefore, damping-like effective field $H_D$ can be calculated using the following equations [29].

$$H_D = -2\frac{H_L}{1-4\xi^2}, \quad (3)$$

$$H_L = \frac{\partial V_{2\omega}/\partial H_X}{\partial^2 V_\omega/\partial H^2_X}, \quad (4)$$

The calculated $H_D$ values are plotted with respect to the applied current as shown in Fig. 3 (d). $H_D$ varies almost linearly with the current, indicating that the effect of Joule heating is negligible. First, we assume that the spin current stems from SHE, and the effective SHA of the multilayers can be extracted using [28-30]:

$$\theta_{SH} = \frac{2|e|M_s t_F H_D}{\hbar |j|}, \qquad (5)$$

where $j$ is the current density and $t_F$ represents the thickness of the Fe layer. The effective SHA value is calculated to be 0.045, which is comparable to the Pt/Co heterostructures, even though the Ru layer has weaker SOC.

To further discuss the origin of the spin currents in the Ru/Fe teterstructures, the harmonic measurements of Cu/Fe and Ru/Co (CoFeB) thin films were carried out as the control experiments. By the same token, to carefully separate ANE and SSE contributions from the measured $V_{2\omega}$, we carried out a series of measurements of field dependent first harmonic and second harmonic Hall voltages with $\varphi$ rotations under various external fields as shown in Fig. 4. According to the theory of Luo *et al*, the the harmonic signals $V_\omega$ and $V_{2\omega}$ can be simplified as:

$$V_\omega = \frac{1}{2} \frac{V_{AHE}}{H_K + H_{ext}} H_Z, \qquad (6)$$

$$V_{2\omega} = (\frac{1}{4} \frac{V_{AHE}}{H_K + H_{ext}} H_D - V_{ANE}) \sin\varphi + \frac{1}{2} \frac{V_{PHE}}{H_{ext}} H_F \cos 2\varphi \sin\varphi, \qquad (7)$$

where $V_{AHE}$, $V_{PHE}$ and $V_{ANE}$ are the anomalous Hall voltage coefficient, planar Hall voltage coefficient and thermo-electric voltage, respectively. $H_D$ and $H_F$ are the damping-like and field-like torque. In this paper, we define $V_1 = V_{AHE} H_D / [4(H_K + H_{ext})] - V_{ANE}$ and $V_2 = V_{PHE} H_F / (2 H_{ext})$. On one hand, for Cu/Fe heterostructure, $V_{2\omega}$ shows a negligible angle dependence. The negligible $V_{2\omega}$ signals could be ascribed to the small SHE of Cu. Considering the similar effective SOC in Ru and Cu, the large SHA in Ru/Fe heterostructures could not stem from the SHE of Ru. The result is consistent with the finding of spin Hall magnetoresistance measurement (see details in supplementary S4). Furthermore, the SHE of Ru should be independent with the FMs. Quite the contrary, the $V_{2\omega}$ signals for Ru/CoFeB heterostructures are much larger than Co. Then, we can obtain the $H_D$ and $H_F$ in Ru/CoFeB heterostructures from the slope of the linear fit as shown in Fig. 4 (d), respectively. Similarly with Ru/Fe, if we also assume that the spin current induced by the SHE of Ru, the SHA of the Ru/CoFeB is calculated to be 0.066 according to the equation (5). These results indicate that the Ru/FM interface strongly influenced the spin current generation. The conclusion can also be confirmed by Ru thickness dependent of $V_{2\omega}$ signals measurements (see details in supplementary S5).

One possible explanation for the spin current generation in Ru/Fe heterostructures is the interfacial spin-orbit scattering [20]. The possible geometric relations between the flow of electrons and accumulated spins in the multilayers are illustrated in Fig. 5(a) and (b), respectively. If one layer is ferromagnetic, the imbalance of majority and minority carriers only arises in nonequilibrium. The interfacial potential has the form [20]:

$$V(r) = \frac{\hbar^2 k_F}{m} \delta(z)[u_0 + u_R \sigma \cdot (\mathbf{k} \times \mathbf{Z})], \quad (8)$$

where $u_0$ is a spin-independent barrier, $u_R$ is the scaled Rashba parameter, $k_F$ is the Fermi momentum, and $\mathbf{k}$ is a unit vector pointing along the incident momentum. Therefore, for normally NM/FM interface, the amplitudes of interface-induced spin polarized current depend on the spin-dependence interface scattering [20]:

$$t^{\pm}(k) = \frac{i k_z k_F}{i k_z k_F - [u_0 \pm u_{eff}(k)]}, \quad (9)$$

where the + (−) labels spins parallel (antiparallel) to the spin-orbit fields. This equation indicates that the spins which are parallel to the spin-orbit fields are easier to transmit the interfaces. Accordingly, even if the interfacical SOC is weak, it might induce an interfacial spin-orbit field. In this case, the reflected and transmitted carriers are spin polarized. Meanwhile, if we have determined the directions of the incident momentum of electrons and the interface normal vectors, the directions of interfacial spin-orbit fields depend on the Rashba parameters. Therefore, according to the schematic diagram as shown in Fig. 5(a), the direction of transmitted spin current flowing into Ru layers is consistent with Rashba parameter and the reflected spin current has opposite direction. This could be the reason of the opposite current-induced magnetization switching behaviors as compared with Pt/Co systems in our preview work which has the same Rashba parameter sign with the Ru/Fe interfaces [28]. Meanwhile, for the Ru/Co interface, it will absorb the minority spin and reflect the majority spin [31]. Therefore, the negligible $V_{2\omega}$ signals in Ru/Co heterostructures as shown in Fig. 4(c) are ascribed to the strong competition that exists at the Ru/Co interface. Besides the spin-orbit filtering current, the spin polarized current will experience spin-orbit procession if the interface exists an interfacial spin-orbit field, as shown in Fig 5(b), and the direction of spin polarization will change to the direction of $\bm{m} \times \bm{\sigma}$. It would reasonably explain the $V_{2\omega}$ signals in Pt/Fe heterostructures with the in-plane magnetic field sweeping along the current flow direction. Meanwhile, as reported in Ref [21], the in-plane electric field (*E//X*) creates non-equilibrium carriers that are anisotropic in momentum space and differ between the FM and NM, which depends on the electrical conductivities. Therefore, the weak interfacial spin-orbit scattering in Cu/Fe heterostructures can be ascribed to the high electrical conductivity of Cu, leading to a small amount of incident carries flowing to the interface and the reflected spin polarized current also decreases correspondingly. More discussion about the influence of electrical conductivity can be found in supplementary S4.

V. P. Amin *et al* [20] have calculated the spin Hall conductivities induced by interfacial spin-orbit scattering based on *ab initio* band structures. Because of the reduced symmetry of the interface, the interface-generated spin current flowing out of plane can be written as [20]:

$$\boldsymbol{j} = j_f\boldsymbol{\sigma} + j_p\boldsymbol{m}\times\boldsymbol{\sigma} + j_m\boldsymbol{\sigma}\times(\boldsymbol{m}\times\boldsymbol{\sigma}), \quad (10)$$

where the three-vector $\boldsymbol{j}$ points along the spin polarization direction. At ferromagnet-nonmagnet interfaces, $j_m$ vanishes when $\boldsymbol{m}$ points in plane or out of plane. However, unfortunately, the magnitude of spin current induced by interfacial spin-orbit scattering with current is absent. At the same time, fortunately, the spin torques induced by the first and second part of equation could be considered as damping-like torque and field-like torque, respectively. Therefore, if we ignore the third part, we could use a simple zero-temperature macrospin model to explain how the current rotates and switches the magnetic orientation $\boldsymbol{m}$ of a perpendicularly magnetized layer and the magnitude of spin current with incident current. We consider a Ru/Fe bilayer in the $XY$ plane with a thickness $t_F$ of the Fe layer, the constant magnetization $M_S$, and the thickness d of the bottom Ru layer. For positive current (electrons flowing in the direction $X$), the interface-induced reflection spin current has spin moments pointing in the direction $Y$ (spin angular momentum along direction $Y$) and flows upward in the direction $Z$. The reflected spin current would be absorbed by FM layers and act as spin transfer torque $\boldsymbol{\tau}_D = \tau_D^0(\boldsymbol{m}\times(\boldsymbol{m}\times\boldsymbol{\sigma}))$, which is oriented along the direction $Y$. We analyze the case with an applied magnetic field $\boldsymbol{H}_{ext} = H_X\boldsymbol{X} + 0\boldsymbol{Y} + 0\boldsymbol{Z}$ (the model is generalizable to arbitrary directions). In addition to the spin torque, we must also take into account the torques (per unit moment) due to the external magnetic field, $\boldsymbol{\tau}_{ext} = -\boldsymbol{m}\times\boldsymbol{H}_{ext}$, and the perpendicular anisotropy field, $\boldsymbol{\tau}_{an} = -\boldsymbol{m}\times\boldsymbol{H}_{an} = -\boldsymbol{m}\times(H_{an}^0 m_z\boldsymbol{Z})$. The equilibrium orientations of $\boldsymbol{m}$ satisfy the condition $\boldsymbol{\tau}_{tol} = \boldsymbol{\tau}_D + \boldsymbol{\tau}_{ext} + \boldsymbol{\tau}_{an} = 0$. We use macrospin simulations with the equation describing the motion $\frac{1}{|\gamma|}\frac{dm}{dt} = \tau_{tol} + \frac{\alpha}{|\gamma|}m \times \frac{dm}{dt}$ with $\alpha > 0$ to distinguish stable from unstable equilibria. If we ignore the thermoelectric effects and the SOT is not strong enough, *m* can be proven to remain in the *YZ* plane. In this case, all torques lie on the *X* axis, and the torque balance equations take the simple form of [4]:

$$\tau_{tol} = x \cdot (\tau_D + \tau_{ext} + \tau_{an}) = \tau_D^0 + H_{ext}\sin\theta + H_{an}^0\sin\theta\cos\theta = 0, \quad (11)$$

With $\theta$ defined as in Fig. 1(a). Parameters $H_{an}^0$ = 1000 Oe, $H_{ext}$ = 150 Oe and $\alpha$ = 0.01 were brought into Equation. Fig. 5(c) shows the magnetic hysteresis curves predicted by this macrospin model for fixed in-plane magnetic fields. As shown in Fig. 3(a), the sign of the hysteresis reverses when the in-plane field component is reversed. At same time, the magnetization switching trajectories under a typical values of spin torque that were calculated with a fixed assistant external field are shown in Fig. 5(d). Under the relatively large damping-like torque condition, the magnetic moment quickly rotates to the opposite hemisphere. The experimental results can be well reproduced by simulation. For Ru/Fe heterostructures, under the critical switching current density $J_c = 1 \times 10^8$ A/cm$^2$, the effective field due to interfacial spin-orbit scattering is $\tau_D^0 = 0.4 H_{an}^0 \approx 400$ Oe in that we ignore the heating effects influence for $H_{an}^0$. Therefore, the ration of $\tau_D^0 / J_c$ is approximate 4 Oe/10$^6$A·cm$^2$. However, there still a problem needs to be further clarified. The real effective field for current induced magnetization switching in our experiment measurements is smaller than 400 Oe because the $H_{an}^0$ decreases significantly as increasing *I* with nonnegligible heating effects.

Recent studies have shown that an additional current dependent unidirectional magnetoresistance (UMR) emerges in NM/FM bilayers due to either the SHE or ISGE

[32]. Unlike the most common magnetoresistive effects, such as the anisotropic magnetoresistance *et al.,* the UMR is a nonlinear effect that is against Onsager reciprocity, being odd under either magnetization or current reversal. Different mechanisms can give rise to UMR in NM/FM systems, even when considering a single source of spin accumulation such as the SHE [33-35]. The UMR provides fundamental insight into the transport properties of spin-orbit coupled systems, including bulk crystals. Here, we have also found the UMR in Pt/Fe heterostructures as shown in Figs. 6(a) and (b). However, the UMR in Ru/Fe heterostructures is absent as shown in Figs. 6(c) and (d), which indicates that the spin current induced by interfacial spin dependence scattering could not effectively accumulate at the Ru/Fe interface as shown in the schematic diagram in Fig. 5(a). Therefore, the UMR could be an effective tool to estimate the origin of spin current. Differently with the report in Ref [21], we have not observed field-free current induced switching in Ru/Fe heterostructures. The possible reason is that, according to the schematic diagram as shown in Fig. 5(a) and (b), the interface-generated spin polarized current obviously relies on the direction of net electron current across the NM/FM interfaces. Therefore, this spin component along the $Z$ direction via interfacial spin procession flowing into Ru layer could not induce field free switching of Fe layer.

In conclusion, we have demonstrated a novel spin current generator through the SOT measurements in Ru/Fe heterostructures that the non-negligible interface-generated spin currents are widely presented in NM/FM heterostructures. Although the effective SHA of Ru in Ru/Fe heterostructures is smaller as compared with that in Ta- or W-based heterostructures, the SOT-induced magnetization switching can still be achieved. Through non-adiabatic harmonic Hall voltage measurements and macrospin simulation, the effective fields induced by Ru/Fe are comparable with that generated by SHE of Pt. Finally, based on the Cu/Fe control experiments, the spin current induced by interface strongly depends on the electrical conductivities. These results establish new opportunities for fundamental studies of spin dynamics and transport in ferromagnetic systems, and a new pathway for efficiently generating strong spin currents for applications.

**Acknowledgements:** This work was partially supported by the National Science Foundation of China (Grant Nos. 51971027, 51927802, 51971023, 51731003, 51671019, 51602022, 61674013, 51602025), and the Fundamental Research Funds for the Central Universities (FRF-TP-19-001A3).

**Figure 1** (a) Field dependence of the normalized out-of-plane magnetization in Pt/Fe heterostructures. (b) First harmonic Hall voltages as a function of a slightly tilted field ($\theta_H = 5°$). Second harmonic Hall voltages with the in-plane component of the tilted field along the $\varphi_H = 0°$ (c) and the $\varphi_H = 90°$ (d). (e) Field dependence of the polar ($\theta_0$) angle of the magnetization. (f) The comparison between the data in (c) and the fitted one using the model of *Hayashi et al*.

**Figure 2** Field dependence of the out-of-plane magnetization (a) and anomalous hall resistance ($R_H$) (b) in Ru/Fe heterostructures with varying temperatures. The Insets in (a) and (b): the out-of-plane magnetization and $R_H$ curves in small magnetic field at room temperature, respectively.

**Figure 3** (a) $R_H$ - $I$ curves in Ru/Fe heterostructures at $H_X = \pm 150$ Oe. (b) $J_c$ as functions of in-plane bias field $H_X$. (c) The first $V_\omega$ and second $V_{2\omega}$ harmonic (inset) Hall voltages plotted against the small in-plane external fields. (d) Current-induced effective field $H_D$ versus applied AC current in Ru/Fe heterostructures.

**Figure 4** (a) The measurement setup along with the definition of the coordinate systems. (b) The $H_Z$ dependence of $V_\omega$ signals. Inset: the $R_H$ curves in large magnetic field. (c) The angle ($\varphi$) dependence of $V_{2\omega}$ signals at $H_{ext} = 1000$ Oe. (d) The $V_1$ and $V_2$ under various external fields.

**Figure 5** (a) and (b) Illustrations of the geometric relation between the flow of electrons, accumulated spins. (c) and (d) Predictions for current-induced magnetic switching and the tracks of magnetic moments within the macrospin model. The green solid line and the red solid line represent the initial and final positions of magnetization, respectively.

**Figure 6** (a) and (b) Unidirectional magnetoresistance measurements in Pt/Fe heterostructures. (c) and (d) Unidirectional magnetoresistance measurements in Ru/Fe heterostructures. All the signals are measured with rotating the external field in the *ZY* plane.

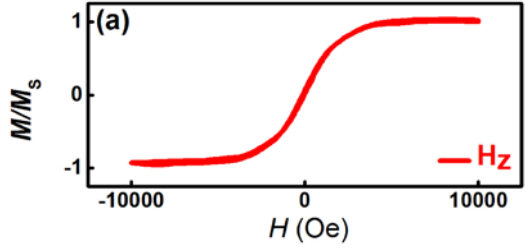
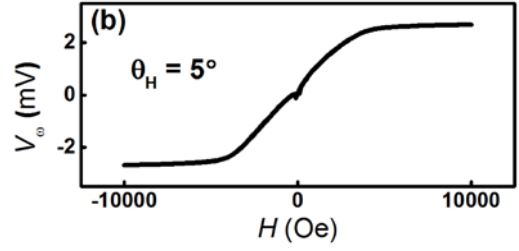
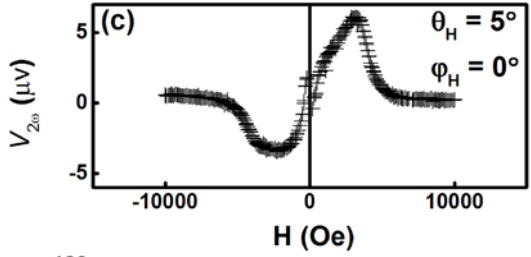
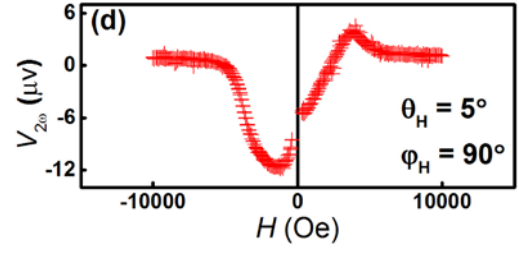
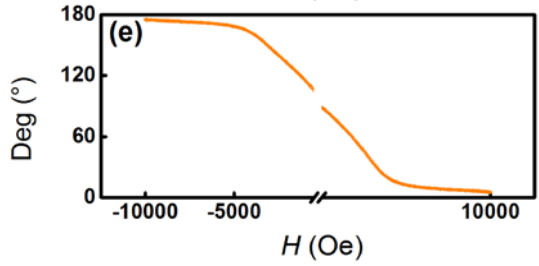
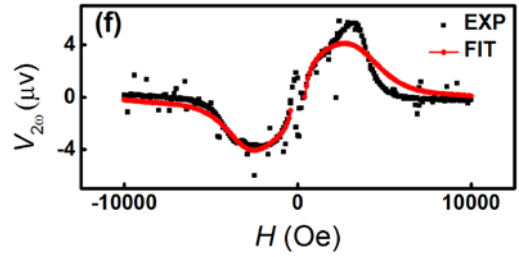
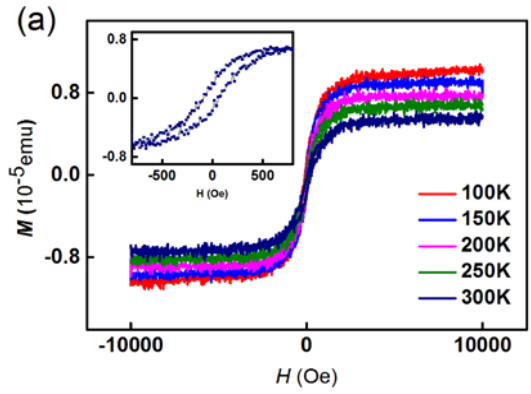
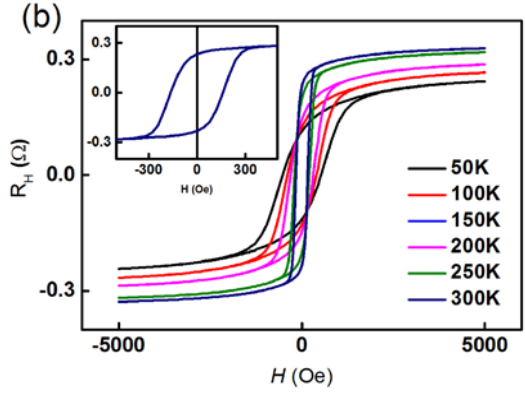

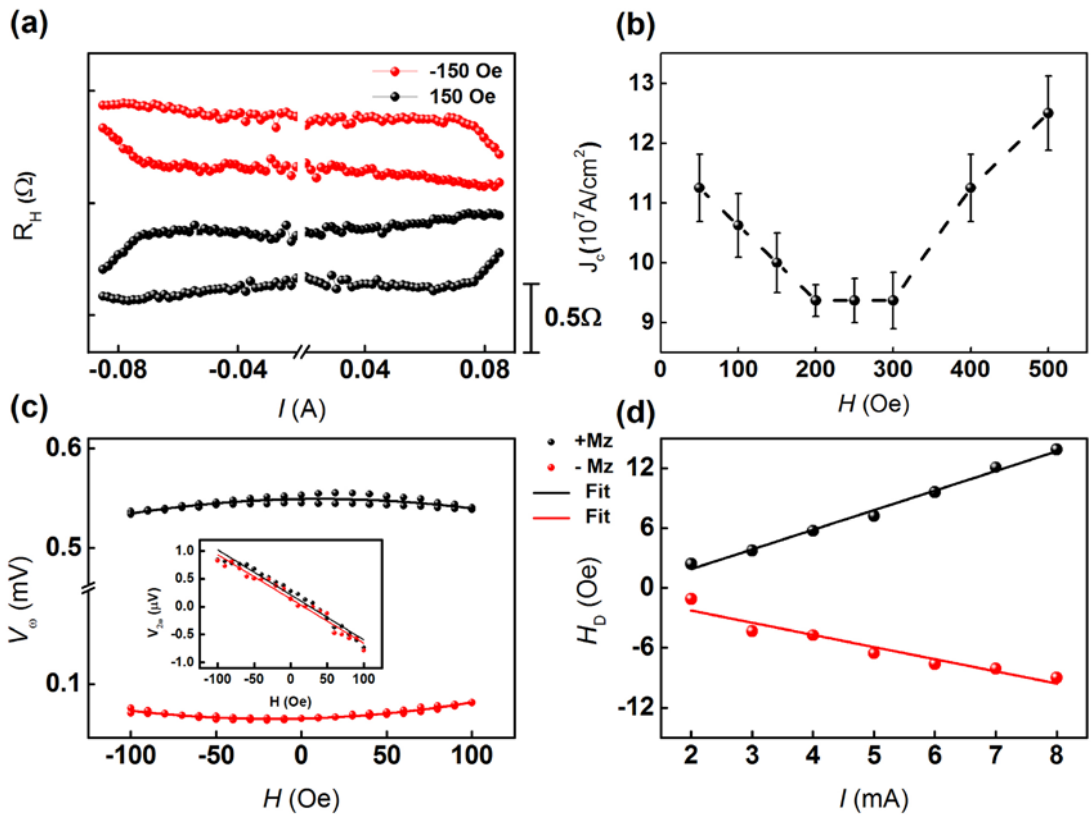

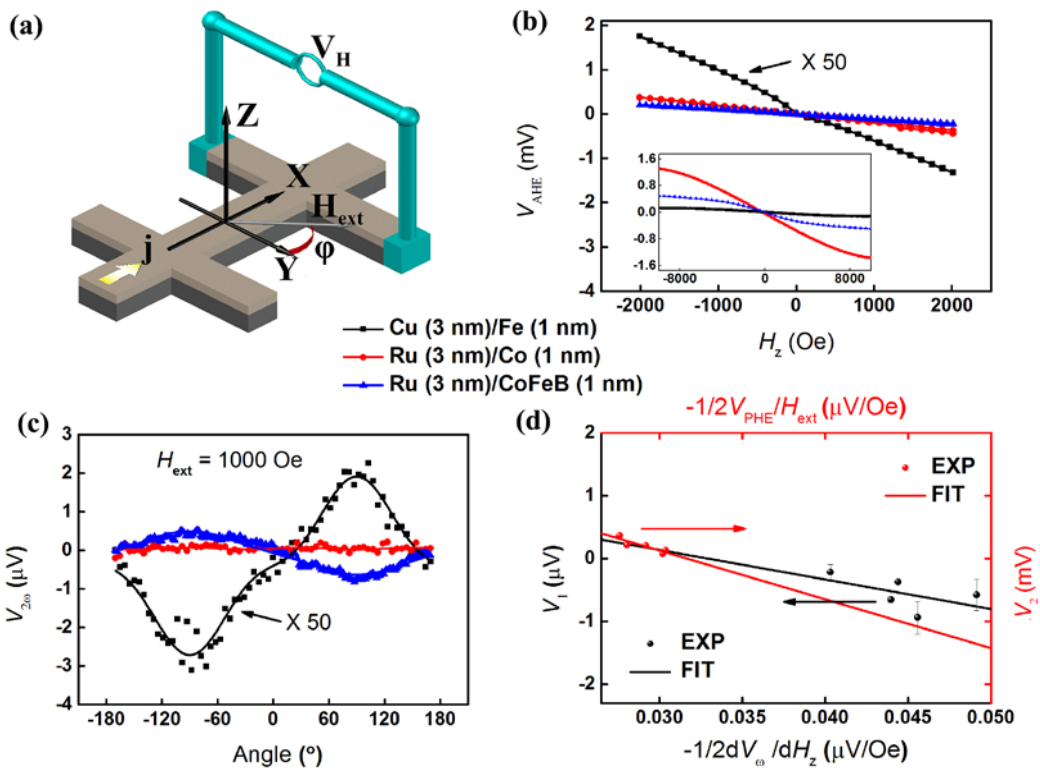

**(a) Spin-orbit filtering** **(b) Spin-orbit precession**

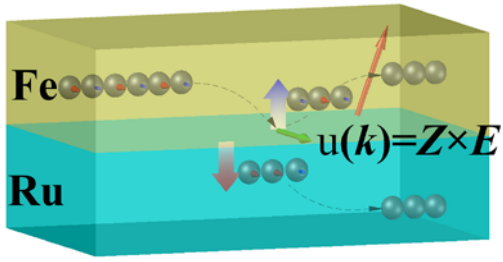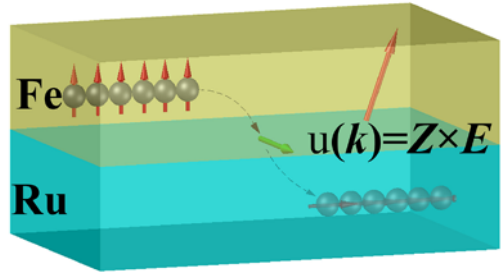

**(c)** **(d)**

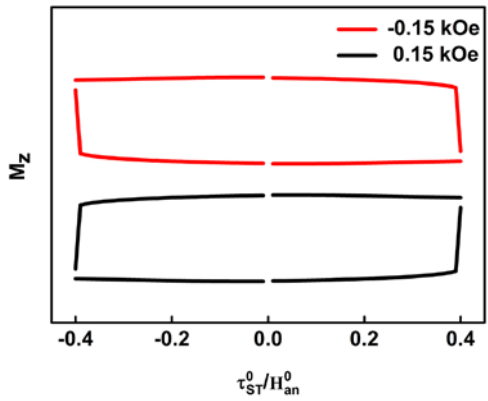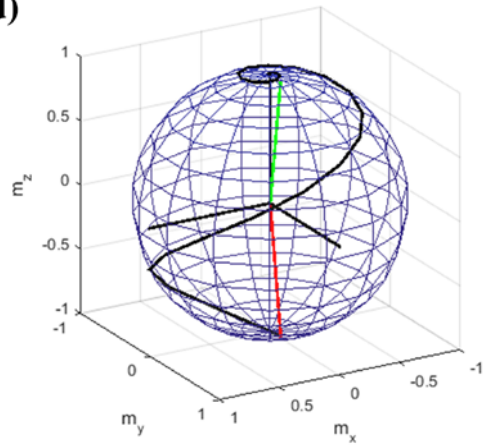

**Pt/Fe** **Ru/Fe**

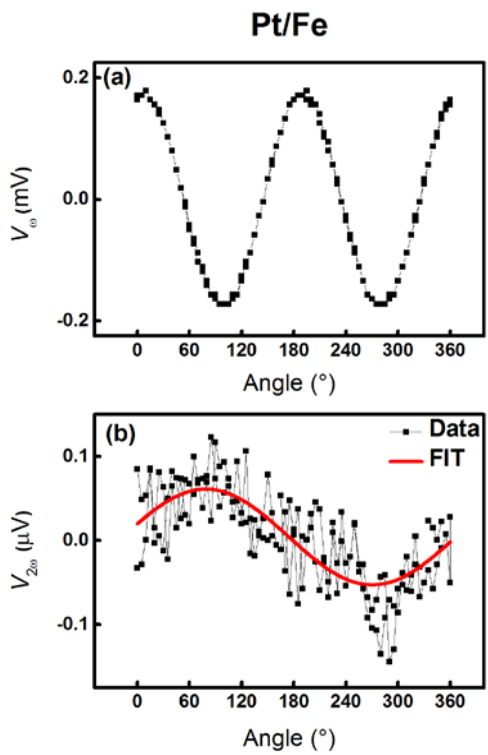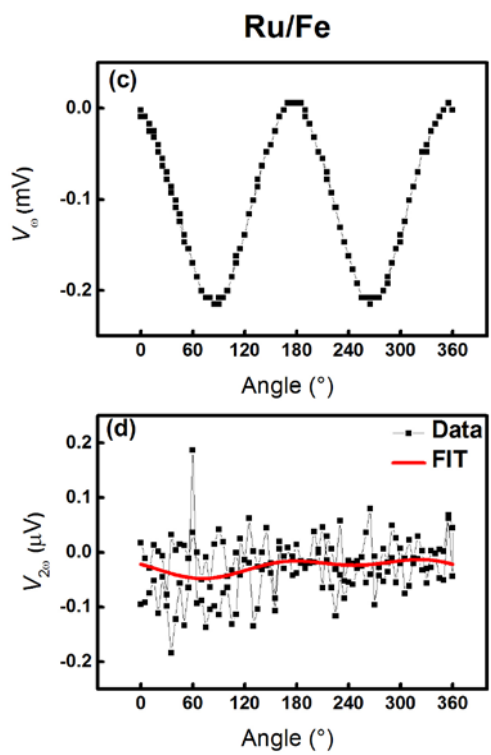